\documentstyle[12pt,aasms4,epsf]{article}
 
\begin{document}

\title{ A rigorous reanalysis of the IRAS variable population: \\
	scale lengths, asymmetries, and microlensing }

\author{Sergei Nikolaev \& Martin D. Weinberg\altaffilmark{1}}

\affil{Department of Physics \& Astronomy \\
	University of Massachusetts, Amherst, MA 01003-4535}

\altaffiltext{1}{Alfred P. Sloan Fellow.}

\abstract{
Previous work reported a bar signature in color-selected IRAS variable
stars.  Here, we estimate the source density of these variables while
consistently accounting for spatial incompleteness in data using a
likelihood approach.  The existence of the bar is confirmed with
shoulder at $a\approx4$ kpc, axis ratio $a:b=2.2$ -- $2.7$ and
position angle of $19^{\circ}\pm1^{\circ}$ degrees.  The ratio of
non-axisymmetric to axisymmetric components gives similar estimate for
the bar size $a=3.3\pm0.1$ kpc and position angle $\phi_0=
24^{\circ}\pm2^{\circ}$. We estimate a scale length $4.00\pm0.55$ kpc
for the IRAS variable population, suggesting that these stars
represent the old disk population.

We use this density reconstruction to estimate the optical depth to
microlensing for the large-scale bar in the Galactic disk.  We find an
enhancement over an equivalent axisymmetric disk by 30\% but still too
small to account for the MACHO result.  In addition, we predict a
significant asymmetry at positive and negative longitudes along lines
of sight through the end of the bar ($|l|\approx30^\circ$) with
optical depths comparable to that in Baade's window.  An infrared
microlensing survey may be a sensitive tool for detecting or
constraining structural asymmetries.

More generally, this is a pilot study for Bayesian star count
analyses.  Bayesian approach allows the assessment of prior
probabilities to the unknown parameters of the model; the resulting
likelihood function is straightforwardly modified to incorporate all
available data.  However, this method requires the evaluation of
multidimensional density functions over the data and optimization of
the function over a parameter space. We address the resulting
computational extremization problem with a hybrid use of a directed
search algorithm which locates the global maximum and the conjugate
gradient method which converges quickly near a likelihood
maximum. Both methods are parallelizable and therefore of potential
use with very large databases.  }


\section{Introduction} \label{sec:intro}

Weinberg (1992, Paper I) identified color-selected variables in the
IRAS Point Source Catalog (PSC) with AGB stars based on color
consistency and the circumstantial sensitivity of the IRAS survey to
long-period variables (cf. Harmon \& Gilmore 1988). These were then
used as rough standard candles to infer a large-scale asymmetry in the
stellar distribution.  The identification of IRAS variables with AGB
stars was strengthened by an in-depth study of a bright subset (Allen,
Kleinmann \& Weinberg 1993).  Carbon-selected AGB stars (carbon stars) have also
proven to be effective tracers (see e.g. Metzger \& Schechter 1994).
Advantages of AGB tracers are reviewed in Weinberg (1994).  In
general, standard candle analyses have the advantage over flux or star
count analyses in providing direct information about the
three-dimensional structure of the Galaxy.  However, uncertainties in
their selection and intrinsic properties may bias any inference and,
especially for the IRAS-selected sample, the census is incomplete.

Paper I described an approach to large-scale Galactic structure using
a star count analysis which allows the information to be reconstructed
and possibly corrected in
the observer's coordinate system before translating to a
Galactocentric system.  Unfortunately, this translation approach is
only natural if the coverage is complete and suffered in application
to the IRAS sample because of spatial gaps due to an incomplete second
full-sky epoch.  Here, we present the results of a different approach
to the problem: the direct density estimation by maximum likelihood.
A Bayesian density estimation has the advantage of directly incorporating
selection effects and missing data.

The number of ongoing surveys that bear on Galactic structure---SDSS,
2MASS, DENIS---which at various stages will have surveyed parts of the
sky is a second motivation for this study; there is a need for a
systematic method suited to inferential studies using possibly
incomplete data from many wave bands.  Recent analyses 
(e.g.  Bahcall \& Soneira 1980 in the optical;
Wainscoat et al. 1992 in the infrared) have modeled the
Galactic components with standard profiles and structural parameters
chosen to provide a match to star count data.  To
explore the structural parameters themselves, we propose a Bayesian
density estimation technique to treat data from scattered fields
during the survey and to easily incorporate data from wave bands.
Conceptually, this approach is midway between a classical inversion
and modeling.  

The first part of the paper describes and characterizes the
method.  More specifically, \S\ref{sec:iras}
reviews the IRAS selection procedure described in Paper I and motivates the
approach.  The new
analysis based on statistical density estimation is presented
in \S\ref{sec:bayes} and precisely defined in \S\ref{sec:likelihood}.
The second part of the paper
describes Monte-Carlo tests and the results of 
applying the method to the IRAS data
(\S\ref{sec:results}).  We conclude in \S\ref{sec:summary} with a
summary and discussion.

\section{IRAS source selection} \label{sec:iras}

The analysis in Paper I was based on the variables selected in the
IRAS Point Source Catalog (1988) by both color and
$P_{var}$.  Following the source selection procedure described in
Paper I, we selected stars from IRAS Point Source Catalog with
$F_{12}>2$ Jy and variability flag $P_{var}\ge 98\%$. Although the flux
limit reduces the confusion in source identification toward
the center of the Galaxy, it also restricts the sensitivity to distant
sources. The limiting distance to a star ($d$) is estimated using a simple
exponential layer with vertical scale height $h$ and mid-plane extinction
coefficient $K_{12}$:
\begin{equation}
m = M + 5 \lg d - 5 + K_{12}\,h\,(1-e^{-d \sin |b| / h})\,/\sin |b|.
\label{eq:b1}
\end{equation}
For a typical AGB star ($L = 3500 L_{\odot}$, see Appendix A) and
$K_{12}=0.18$ kpc$^{-1}$, the limiting distance in the plane is
$R_{lim}=7$ kpc.  We assume that the extinction is dominated by the
molecular gas, $h=100$ pc and the extincting layer is horizontally
isotropic.  The true extinction toward the inner Galaxy is most likely
dominated by the molecular ring and nuclear region given the molecular
gas distribution.  However, precise estimate of the true distribution
is not available and an horizontally isotropic model will adequately
represent its systematic effect on the photometric distances.

Of the more than 158,000 good flux-quality sources listed in IRAS PSC,
5,736 satisfy both flux limit and variability criteria.  Their spatial
distribution is shown in Figure \ref{fig:sp_distr.1}.  To obtain
variability data, at least two epochs are needed.  Unfortunately,
IRAS' multiple epochs did not have complete sky coverage.  Most of the
coverage (77\% in the galactic plane) was achieved in HCON 2 and HCON
3 separated by roughly 7.5 months on average. The rest of the galactic
plane is poorly sampled (shaded regions in Figure
\ref{fig:sp_distr.1}).  For this analysis, all the data in the poorly
sampled sectors have been excised, reducing the size of the sample to
5,500 stars.

\section{Method overview} \label{sec:bayes}

All of the selection effects but especially data incompleteness
greatly complicate the analysis.  Bayesian techniques are ideally
suited to parameter estimation over data with general but well-defined
selection criteria and underlies both the maximum entropy and maximum
likelihood procedures.  Below, we will parameterize the source density
by an exponentiated orthogonal series with unknown coefficients
$A_{ij}$ and $B_{ij}$ (cf. eq. \ref{eq:d15}). In this context, the
basic theorem of the theory reads:
\begin{equation}
P \, ( \{A_{ij}\} ,\, \{B_{ij}\} \, | \, D, \, I \,) = 
  {{ P \,(\{A_{ij}\} ,\, \{B_{ij}\} \, | \, I \,) \cdot
    P \,( D \, | \, \{A_{ij}\} ,\, \{B_{ij}\} ,\, I \,) } \over
    P \,( D \, | \, I \,)}. \label{eq:d1}
\end{equation}
The probability $P \, ( \{A_{ij}\} ,\, \{B_{ij}\} \, | \, D, \, I \,)$
is the conditional (or {\it posterior}) probability of the
coefficients of the source density provided the data ($D$) and
information ($I$) describing its incompleteness.  The probability
$P\,(\{A_{ij}\} ,\, \{B_{ij}\} \, | \, I \,)$ is the prior probability
(or simply, {\it prior}) of the coefficients provided only the
information.  Following Bretthorst (1990), we assign the prior using
the maximum entropy principle.  In our case it is constant implying
that all coefficient values are equally likely initially.  The
function $P \,( D \, | \, \{A_{ij}\} ,\, \{B_{ij}\} ,\, I \,)$ is the
direct probability which describes the likelihood of data given the
coefficients.  Finally, $P \,( D \, | \, I \,)$ is a normalization
constant which may be omitted provided that the posterior probability
is normalized.

With these definitions, it follows that
\begin{equation}
P \, ( \{A_{ij}\} ,\, \{B_{ij}\} \, | \, D, \, I \,) =  \mbox{Const} \cdot
  P \,( D \, | \, \{A_{ij}\} ,\, \{B_{ij}\} ,\, I \,), \label{eq:d2}
\end{equation}
or in words, the posterior probability is proportional to the
likelihood function.  Therefore, the best estimate of posterior
probability is obtained for the set coefficients which maximize the
likelihood function.

\section{Likelihood function} \label{sec:likelihood}

The likelihood is the joint probability of the observed stars given a
source density.  We may then consider the probability of observing a star with
intrinsic luminosity in the range $( L, L+dL )$ to be detected in the
distance interval $( s, s+ds )$, in the azimuth interval $( l, l+dl )$, in
the galactic latitude interval $( b, b+db )$
and with magnitude in the range $( m, m+dm )$.  Assuming a normal
distribution of intrinsic luminosities $L$ and a normal error
distribution for the apparent magnitudes $m$ this becomes:
\begin{eqnarray}
P_n\,(s,\,l,\,b,\,m,\,L\,|\,\sigma_m,\,\sigma_L,\,K_{12},\,h,\,R_0)\,
s^2\,ds\,\cos b\,db\,dl\,dL\,dm = \nonumber \\ 
C \cdot \Sigma (r,\,\phi,\,z)\,e^{-{(L-\overline L)}^2/2 \sigma_L^2}\,
e^{-{(m-\overline m)}^2/2 \sigma_m^2}\,
s^2\,ds\,\cos b\,db\,dl\,dL\,dm. \label{eq:d3}
\end{eqnarray}
Here $s$, $l$, $b$ are coordinates about the observer's position, $r$,
$\phi$, $z$ are coordinates about the center of the Galaxy, $C$ is the
normalization constant, $\Sigma (r, \phi, z)$ is the source density at
galactocentric radius $R_0$, $\overline L$ and $\sigma_L$ are the mean
intrinsic luminosity and the dispersion of the sample, $\sigma_m$ is
the measurement error in magnitudes and $\overline m = \overline
m\,(s, b)$ is given by equation (\ref{eq:b1}).  Alternatively, we may
replace luminosity by absolute magnitude:
\begin{eqnarray}
P_n\,(s,\,l,\,b,\,m,\,M\,|\,\sigma_m,\,\sigma_M,\,K_{12},\,h,\,R_0)\,
s^2\,ds\,\cos b\,db\,dl\,dM\,dm = \nonumber \\
C \cdot \Sigma (r,\,\phi,\,z)\,e^{-{(M-\overline M)}^2/2 \sigma_M^2}\,
e^{-{(m-\overline m)}^2/2 \sigma_m^2}\,
s^2\,ds\,\cos b\,db\,dl\,dM\,dm,\label{eq:d4}
\end{eqnarray}
where $\overline M$ and $\sigma_M$ correspond to $\overline L$ and
$\sigma_L$. The Gaussian distributions in $L$ or $M$ in the above two
equations can be generalized to an arbitrary luminosity function for
traditional star count applications.  Although we will not give the
general expressions below, the development is parallel.

Since the convolution of two Gaussians is a new Gaussian whose
variance is the sum of the two individual variances
\begin{equation}
\sigma_{m, eff}^2 = \sigma_m^2 + \sigma_M^2, \label{eq:d5}
\end{equation}
equation (\ref{eq:d4}) can be rewritten as
\begin{eqnarray}
P_n\,(s,\,l,\,b,\,m\,|\,\sigma_{m, eff},\,k,\,H,\,R_0)\,
s^2\,ds\,\cos b\,db\,dl\,dm = \nonumber \\
C \cdot \Sigma (r,\,\phi,\,z)\,e^{-{(m-\overline m)}^2/2 \sigma_{m, eff}^2}\,
s^2\,ds\,\cos b\,db\,dl\,dm \label{eq:d6}
\end{eqnarray} 
after integrating over the unmeasured absolute magnitude $M$. 
For notational clarity, we will
omit the subscript ``eff'' and write simply $\sigma_m$.  The constant
$C$ is determined from the normalization condition:
\begin{equation}
C \int _{-\infty} ^{+\infty} e^{-{(m-\overline m)}^2/2 \sigma_m^2}\,dm
\int dl  \int _0 ^{s_{max}(b)}\,s^2\,ds \int _{-{\pi \over 2}} ^{\pi \over 2}
{ \Sigma (r,\,\phi,\,z)\,\cos b\,db} = 1. \label{eq:d7}
\end{equation}
The integration over $l$ runs over entire circle except missing
azimuthal sectors, explicitly accounting for missing data at
particular ranges in azimuth.  The limiting distance $s_{max}$ in the
$l$, $b$ direction incorporates the 2 Jy flux limit.  

In a standard star count analysis no explicit distance information is
provided and $s$ is eliminated from analysis by integration, yielding
\begin{eqnarray}
P_n\,(l,\,b,\,m\,|\,\ldots)\,\cos b\,db\,dl\,dm = \nonumber \\
C \,\int _0 ^{s_{max}(b)} {\Sigma (r,\,\phi,\,z)\,
e^{-{(m-\overline m)}^2/2 \sigma_m^2}\,s^2\,ds} \cos b\,db\,dl\,dm. 
\label{eq:d9}
\end{eqnarray}
For our relatively small sample of IRAS stars, sensitivity to vertical
structure will be poor.  This motivates replacing the general unknown
three-dimensional disk density with a density which depends on radial
position and azimuth alone: $\Sigma (r,\,\phi,\,z) = {\overline
\Sigma} (r,\,\phi)$.

Finally, the joint probability of observing $N$ stars selected from
the IRAS PSC is
\begin{equation} 
L \equiv P_{total} = \prod _{n=1} ^N P_n ( l,b,m | \ldots).\label{eq:d11}
\end{equation}
Expressing the likelihood function in logarithmic form, our desired
solution is the set of parameters which maximize
\begin{equation} 
\log  L = \sum _{n=1} ^N \log P_n ( l,b,m | \ldots).\label{eq:d12}
\end{equation}

This and nearly all star count analyses reduce to standard problem of
density estimation: find the density function $f ( x )$, which
satisfies non-negativity constraint
\begin{equation}
f(x) \ge 0 \label{eq:d13}
\end{equation}
and integral constraint
\begin{equation}
\int f(x)  dx = 1 \label{eq:d14}
\end{equation}
which best describes the observed data distribution.  Both parametric
and non-parametric estimation techniques have been used to solve this
problem (e.g. Silverman 1986; Izenman 1991).  For inhomogeneous
multidimensional data, the positivity constraint is
cumbersome. However, searching for the unknown function $f ( x )$
in the form of an exponentiated orthogonal series (Clutton-Brock
1990), guarantees positivity.  A candidate stellar surface density is:
\begin{equation}
\overline \Sigma (r,\phi) = \exp \biggl\{ \sum_{i=1} ^{i_{max}} \sum_{j=0}
^{j_{max}} \left[ A_{ij}\cos j\phi + B_{ij}\sin j\phi \right]
J_j(k_i^j r) \biggr\} ,\label{eq:d15}
\end{equation}
where $J_j(x)$ is Bessel function of $j^{\hbox{th}}$ order and $k_i^j$
is $i^{\hbox{th}}$ root of Bessel function of $j^{\hbox{th}}$ order
and are chosen to produce a complete orthogonal set over the disk of
radius $R_{max}$. The coefficients $A_{ij}, B_{ij}$ are the parameters
to be determined.  There is no loss of generality in taking the
Fourier-Bessel series although the choice is arbitrary.

\section{Results} \label{sec:results}

\subsection{Sensitivity to incompleteness}

A major advantage of the approach presented here over that in Paper I
is that the significance of inferred structure is robustly
quantified. In particular, we can test the sensitivity of selection
effects to the detection of a bar.  To test the presence of the
coverage gaps, we generated four sample disks of 1,000 stars each
using the source density (\ref{eq:d15}) with
$\sqrt{A_{ij}^2+B_{ij}^2}=1$ for $j=0,2$ and zero otherwise and the
following bar position angles: $0^\circ$, $\pm45^\circ$, and
$90^\circ$.  The root sum square of the coefficients $A_{ij}$ and
$B_{ij}$ represents the strength of $i^{th}$ radial component for the
$j^{th}$ polar harmonic.  Figure \ref{fig:test} shows the restored
strength of a harmonic $\sqrt{A_{ij}^2 + B_{ij}^2}$ as a function of
the position angle of the bar.
Insensitivity of these strengths to bar position angle suggests that
missing azimuths will not obscure the inference of true bar. The
computed values are consistent with the expected value of unity.

Conversely, regions of missing data can produce non-axisymmetric
distortions, and in principle, suggest the existence of a bar in
initially axisymmetric sample.  However, analysis of a simulated
axisymmetric disk ($A_{10}=A_{20}=1$; all others = 0) and the same
azimuthal incompleteness as in the real sample shows that the power in
the non-axisymmetric harmonics is about 3\% of the axisymmetric
contribution.  Together these tests suggest that the misidentification
of a bar relative due to missing azimuthal sectors alone is unlikely.

\subsection{Application to IRAS data}

The formalism developed in \S\ref{sec:likelihood} requires
the distance to galactic center $R_0$, extinction in the
plane $K_{12}$ and average luminosity of the AGB stars $\overline L$.
We adopted $R_0=8.0$ kpc, $K_{12}=0.18$
mag/kpc and $\overline L = 3500 L_{\odot}$. The method can
be straightforwardly modified for complex models (e.g. patchy or non-uniform
extinction), the only limitation
here is the CPU available and sufficient data to attain a satisfactory
measure of confidence.

Choosing the truncation of the series in equation (\ref{eq:d15}) poses
a problem common to many non-parametric density estimations: because
too few terms result in large bias and too many terms increase
variance, $i_{max}$, $j_{max}$ would be best determined by jointly
minimizing the bias and the variance.  However, this approach is
computationally prohibitive due to the integral in
equation (\ref{eq:d9}) and the normalization (\ref{eq:d7}). Therefore,
a heuristic approach was adapted in selecting $i_{max}$, $j_{max}$
based on the increase in the likelihood function when a particular
term or set of terms is added. Significance could be quantified in
terms of the likelihood ratio (Wilks 1962) but we have not done this
here. In addition, the hardware available to us makes it impossible to
sample the parameter space beyond $i_{max}=4$,
$j_{max}=4$. Nevertheless, up to that limit, the space was sampled
thoroughly, with some of the solutions shown in Figure
\ref{fig:dens_all} along with the corresponding offsets of the
likelihood function (the lowest value of likelihood is set to $0$ for
ease in comparison).
Some of the figures feature the ghost peaks due to the absence of data
beyond the galactic center or in missing azimuthal sectors (see
Figs. \ref{fig:sp_distr.1} and \ref{fig:sp_distr.2}). The likelihood
analysis may attempt to place a non-existing source density peak in
that region, provided it will increase the overall score.  We will
pursue penalizing the likelihood function and other procedures for
choosing an alternative prior (dropping the assumption that all
coefficients in (\ref{eq:d15}) are equally likely initially) in future
work.

More importantly, all reconstructions in Figure \ref{fig:dens_all}
imply a jet-like feature in the first quadrant. As in Paper I, the
depth of our sample (estimated to correspond to a mean distance of 7
kpc in the plane) prevents ascertaining whether this feature
corresponds to a bisymmetric bar or is a lopsided distortion.
However, decreasing the flux limit to 1 Jy leads to detection of
similar feature on the far side of the Galaxy, suggesting a real
bar. This motivates a reconstruction with enforced bisymmetry, shown
in Figure \ref{fig:dens_even}.  Here the corresponding prior assigns
zero values to coefficients of odd azimuthal order.
The likelihood value (the origin is the same as in Figure
\ref{fig:dens_all}) has dropped substantially, because the resulting density
lacks data support  beyond the Galactic center. 
In both figures, the bar is well defined and has a similar length and 
position angle.  

To quantify the strength and position angle of the bar, we fitted the
isodensity contours ($i_{max}=j_{max}=4$) by ellipses.  The logarithm
of a suitable likelihood function for estimating the semi-major axes,
eccentricity and position angle is
\begin{equation}
\log L = \sum _{i=1} ^M {\biggl[ \Sigma_{rec}(r_i, \phi_i)-C \biggr]}^2,
\label{eq:d17}
\end{equation}
where $\Sigma_{rec}(r, \phi)$ is the reconstructed density function and
$C$ is isodensity level. The summation runs over equally spaced points
on ellipse.  For a given ellipse, a grid of semimajor axis values are
specified and the surface density $C$, position angle $\phi_0$ and
eccentricity $e$ which maximizes $\log L$ are found.  The results
are presented in Figures \ref{fig:levels} and \ref{fig:angles}.

Figure \ref{fig:levels} indicates that the density profile drops to
half of its central value at about 4 kpc.  The half-length would then
be about 4 kpc, in good agreement with the value obtained in Paper I.
If we take this value as the size of the major axis of the bar, then
the axis ratio varies from 2.2 in the central regions to 2.7 in the
outer regions of the bar.  The value of the position angle for the
entire extent of the bar (out to 4 kpc) is $\approx 19^{\circ}$. The
accuracy of the position angle determination can be quantified in
terms of confidence interval, making use of the fact that in the limit
of large number of sources $N$, the likelihood in $n$ dimensions is
distributed as $\chi^2/2$ with $n$ degrees of freedom (e.g. Lehmann
1959). We analyzed the likelihood as the function of a single variable
-- orientation angle of the bar in the plane. The analysis gives the
uncertainty of $1^{\circ}$ at $3\sigma$ level.

Another way to determine the parameters of the bar is to look at the
map of the ratio of non-axisymmetric to axisymmetric components of the
density. The ratio displays two peaks at $3.3\pm0.1$ kpc located on
the opposite sides from the center, the line connecting them has the
position angle of $\sim24^{\circ}\pm2^{\circ}$.  The peak ratio, the
relative strength of the bar, is $0.73$.  This implies the existence
of a strong bar in the intermediate age population responsible for the
AGB stars.

\subsection{Disk scale length}

Having calculated the source density, we are in a position to
characterize the parent population of the IRAS variables.  In Paper I,
we assumed that these variables represented a disk population based on
their flux distribution but several colleagues have suggested in
discussion that the IRAS variables are more likely to be bulge stars.
Here, we determine the scale length of the population in the Galactic
plane.  For comparison, we fit our reconstruction by an oblate
spheroid model (G0 bulge model from the DIRBE study by Dwek et
al. 1995):
\begin{equation} 
\Sigma_{G0} ( x,  y ) = \Sigma_0 e^{-0.5 r^2},
\label{eq:d18}
\end{equation}
with $r^2 = ( x^2 + y^2 ) / r_0^2$.  The scale length $r_0$ is found
by minimizing the following cost function while simultaneously
satisfying the overall normalization constraint for $\Sigma_{G0}$
(eq. \ref{eq:d14}):
\begin{equation} 
\hbox{cost} = \int d^2 r  {\biggl[ \Sigma_{rec}-\Sigma_{G0} \biggr]}^2.
\label{eq:d19}
\end{equation}
To estimate the value of $r_0$, we used the covariance matrix from
the likelihood 
analysis used to determine $\Sigma_{rec}$ to make 5000 Monte Carlo
realizations of the source density.  The ensemble of realizations,
then, have $\Sigma_{rec}$ as their mean.  For each realization, we
found $r_0$ by minimizing the cost function (\ref{eq:d19}) and the
resulting distribution of scale lengths is shown in Figure
\ref{fig:dwek}.  Our result $r_0 = 4.00\pm0.55$ kpc indicates that the
IRAS variables have the scale length of the old disk population. This
value is in good agreement with the scale length $4.5$ kpc reported by
Habing (1988), derived from analysis of a color-selected IRAS sample.
Dwek's value obtained by analyzing bulge emission was
$r_0=0.91\pm0.01$ kpc.  The factor of $4$ difference between the scale
lengths suggests that the IRAS bar and the bulge-bar belong to
distinct populations.

\subsection{Optical depth due to microlensing}

Originally proposed as a test for dark matter in the Milky Way halo
(Paczy\'nski 1986), gravitational microlensing was later shown (Griest
et al. 1991; Paczy\'nski 1991) to be potentially useful for extracting
information about the inner regions of our Galaxy. Three groups (OGLE,
MACHO and EROS) are monitoring stars in the Galactic bulge for
gravitational microlensing and have found higher event rates
than most theoretical estimates.  Udalski et al. (1994) derived
lensing optical depth $\tau = (3.3 \pm 1.2) \times 10^{-6}$ toward the
Baade's window ($l = 1^{\circ}, b = -3.9^{\circ}$) based on the
analysis of the OGLE data, and MACHO group reported $\tau =
3.9^{+1.8}_{-1.2} \times 10^{-6}$ (Alcock et al. 1995a) estimated from
the sample of clump giants, while theoretical estimates give optical
depths in the range $0.5 - 2.0 \times 10^{-6}$ (e.g. Alcock et
al. 1995a; Evans 1994).  Following Paczy\'nski's et al. (1994) suggestion that
a bar with a small inclination angle could enhance the optical depth,
Zhao et al. (1995) have developed a detailed bar model and found $\tau
= (2.2 \pm 0.5) \times 10^{-6}$.
Here, we estimate the optical depth using our density
reconstruction, $\Sigma_{rec}$, assuming that our AGB sample represents
the entire stellar disk.

The lensing optical depth is defined as the probability of any of the
sources being lensed with magnification factor $A > 1.34$, with
\begin {equation}
A = {u^2+2 \over u \sqrt{u^2+4}}, \qquad  u \equiv {r \over R_E}
\label {eq:d20}
\end {equation}
(Refsdal 1964), where $r$ is the distance between the projected position 
of the source and the lensing mass, $R_E$ is the radius of Einstein ring.
Kiraga \& Paczy\'nski (1994) derived
\begin {equation}
\tau = {4 \pi G \over c^2}\,\, 
{\int _0 ^{\infty} \left[ \int _0 ^{D_s} \rho \, {D_d(D_s-D_d) \over D_s} \,\, dD_d \right] 
\rho \, D_s^{2+2\beta}\, dD_s \over \int _0 ^{\infty} \rho \, D_s^{2+2\beta}\, dD_s},
\label {eq:d21}
\end {equation}
where $D_s$ is the distance to the sources, $D_d$ is the distance to
the deflectors and the free parameter $\beta$ accounts for
detectablity of sources in a flux-limited survey. The reasonable
range is $-3 \le \beta \le -1$ and we take $\beta = -1$ following
Evans (1994) and Kiraga \& Paczy\'nski (1994). 
The density $\rho = \rho_{bulge}
+ \rho_{disk}$, with $\rho_{bulge}$ given by equation (1) of Kent
(1992), and
\begin {equation}
\rho_{disk} = C \, \Sigma_{44} (r, \phi) \, e^{-|z|/h}, 
\label {eq:d22}
\end {equation}
where $\Sigma_{44}$ is the surface density of our $i=4, j=4$ model (\ref{eq:d15}) and 
$h=0.325$ kpc is the scale height. 
We explored two possible normalization prescriptions: (1) Assign a
local column density of $\sim 50\, M_{\odot}\, pc^{-2}$ (``canonical
disk'' following Kuijken \& Gilmore 1989; Gould 1990). The mass of the
disk in this case is $M_{disk} = 1.95 \times 10^{10} M_{\odot}$. 
(2) Assign the total disk mass of $M = 6 \times 10^{10} M_{\odot}$
(Bahcall \& Soneira 1980).  The second normalization gives local
column density of approximately $100\, M_{\odot}\, pc^{-2}$ (``maximal
disk'' of Alcock et al. 1995b).  We prefer the latter here because the
optical depth estimate depends on the global mass distribution rather
than the local density.  In addition, there are some indications that
the variation of the column density with galactic longitude may be
quite significant -- a factor of $2-3$ (Rix \& Zaritsky 1995; Gnedin,
Goodman \& Frei 1995).  The mass of the bulge is $M_{bulge} = 1.65
\times 10^{10} M_{\odot}$.

For the canonical disk case, the total lensing optical depth at
Baade's window is $1.1 \times 10^{-6}$, and both bulge and disk lenses
contribute $50\%$ to that number. Most of the optical depth (76\%) is
due to lensing of bulge sources. If the disk is maximal, optical depth
is $1.6 \times 10^{-6}$. Disk lenses now account for $1.1 \times 10^{-6}$
(68\% of the total optical depth) and the contribution by bulge sources
still dominates (59\%).
For both scenarios, optical depth is a function of the
orientation of the bar.  We investigate the enhancement produced by
the bar over axisymmetric models of the disk $\rho\propto e^{-r/R} \,
e^{-|z|/h}$, where $R = 3.5$ kpc for fixed disk mass.  Figure
\ref{fig:baadetau} displays the ratio of optical depths of
non-axisymmetric to axisymmetric disk models as a function of the
position angle of the bar for both normalization scenarios.
The difference between the two curves illustrates the role of the disk
in lensing.  The largest enhancement of approximately 30\% obtains
when the bar is aligned along the line of sight as expected. The ratio
of optical depths decreases gradually when the bar is in the first
Galactic quadrant, with $\ge 20\%$ enhancement out to $\phi_0 =
50^{\circ}$.

Current generation optical-band lensing surveys have concentrated on
low-extinction bulge-centered windows to maximize the lensing event
rate.  An infrared-band lensing microlensing survey would be less
constrained by extinction and therefore more efficient probe of the
overall structure of the Galaxy.  In particular, any bar which is not
perfectly aligned along the Sun--Galactic Center axis will produce an
asymmetry in the optical depth.  We describe this asymmetry by the ratio
of the difference in optical depths at positive and negative longitude
to their arithmetic mean. This ratio is shown in Figure \ref{fig:reldiff} for
our model (cf. eqns. \ref{eq:d21} and \ref{eq:d22}).  Comparison with
the Bahcall \& Soneira model (1980) suggests that $\beta\approx-1$ is
a fair approximation of the high-luminosity end of the disk luminosity
function.  Therefore, equation (\ref{eq:d21}) also applies at large
$|l|$ where both lenses and sources are disk members.  The large 40\%
asymmetry about $|l|\approx30^{\circ}$ is due to a local increase in
the surface density at negative longitudes close to the observer
(Figure \ref{fig:dens_even}).  More important than the details of
asymmetry is the suggestion that a pencil-beam microlensing survey in
the infrared would be sensitive to global asymmetries in the stellar
disk component. Confusion is not a limitation at $b=0^\circ$ for larger
values of $| l |$ and the optical depth has a magnitude similar to Baade's
window.

\section{Summary and discussion} \label{sec:summary}

This paper explores a model-independent Bayesian estimation of the
stellar density from star counts, rigorously accounting for incomplete
data.  The general approach can incorporate multiple colors and even
different databases.  The usual high dimensionality and topological
complexity of the posterior distribution, however, complicates both
optimization algorithms and subsequent moment analyses.  We propose
here a hybrid downhill plus directed-search Monte Carlo algorithm; the
former speeds convergence and the latter facilitates the location of
the global extremum.  Other similar and potentially more efficient
techniques which can bypass the extremization step altogether (such as
general Markov Chain Monte Carlo) are worth careful consideration.

Application of the technique to the variability-selected sample
described in Weinberg (1992), assumed to be AGB stars, confirms the
presence of a strong non-axisymmetric feature in the first Galactic
quadrant. By imposing bisymmetry on the source density, clear
signature of a bar is obtained.  The size and shape of density
isophotes suggests a bar semi-major axis of approximately 4 kpc and
position angle of $\phi_0 = 18^\circ \pm 2^\circ$ at the outer edge of
the bar.  The analysis of the scale length for the AGB candidate
distribution gives $r_0=4.00\pm0.55$ kpc, indicating that these
objects are part of the old disk population.

Finally, we use our estimate for non-axisymmetric Galactic disk to
explore the dependence of optical depth to gravitational microlensing
by bulge and disk stars.  The disk bar does enhance the optical depth
$\tau$ towards Baade's window by roughly 30\% but the overall value is
still roughly a factor of two below the MACHO result $\tau =
3.9^{+1.8}_{-1.2} \times 10^{-6}$.  Of interest for future
microlensing surveys is the finding that our inferred large-scale bar
will produce a significant asymmetry in $\tau$ at positive and
negative longitudes beyond the bulge.  The peak asymmetry for our
model occurs at $|l|=30^\circ$ and at $b=0$ we predict similar values
of $\tau$ to the Baade's window field.  Such a survey might best be
carried out in the infrared to take advantage of the low interstellar
extinction and colors of the late-type giants.  At
$|l|\gtrsim30^\circ$, confusion should not be a limitation at
$b=0^\circ$.

\acknowledgements

We thank Steve Price and Mike Skrutskie for comments.  This work was
supported in part by NASA grant NAG 5-1999 and the Alfred P. Sloan
Foundation.

\appendix

\section{Luminosities of AGB stars}

The luminosities of AGB variables and the inference of their
progenitor masses plays a role in constraining the stellar evolution
history of the Galaxy and has received some attention.  Investigations
based on theoretical approach (Iben \& Renzini 1983) and observations
of sources close to the Galactic center (Jones \& Hyland 1986) placed
the luminosities somewhere between a few $\times 10^3 L_{\odot}$ and $6 \times 10^4
L_{\odot}$. Van der Veen \& Habing (1990) revised the results of Jones
\& Hyland based on the analysis of a larger sample of OH/IR stars and
found that the luminosities are in the range $10^3$ -- $10^4
L_{\odot}$ with the peak of the distribution about $5,000 - 5,500
L_{\odot}$. They suggested the variability of the sources ($\Delta m
\le 2^m$) and possible selection effects as main reasons for higher
limits of Jones \& Hyland. They also noted that as many as 20\% of the
stars may be in the low-luminosity tail of the distribution but only
2\% or fewer can exceed the upper limit.  Kastner et al. (1993)
obtained kinematic luminosities based on the radial velocities of
circumstellar envelopes with respect to the LSR and distances derived
from the Galactic rotation curve. They found the range of $1.3 \times
10^4$ -- $2 \times 10^4 L_{\odot}$ with average uncertainty of factor
of 2.  The theoretical estimate was recently revised by Groenewegen et
al. (1995) who obtained luminosity functions for carbon and
oxygen-rich stars based on the synthetic evolution. They found a mean
luminosity for Galactic carbon and oxygen-rich AGB stars to be $7050
L_{\odot}$ and $3450 L_{\odot}$, respectively. They stated ``the
luminosity of a typical Galactic AGB star is in any case less than the
$10^4 L_{\odot}$ often assumed''.  Habing (1988) reported the average
luminosity of $4000 L_{\odot}$ for a color selected sample from IRAS
PSC catalog.  Finally, analysis of a sample of oxygen Miras using P--L
relation established on the observations of LMC Miras (Feast et
al. 1989), places their average luminosity at $\overline L =
3900\pm450 L_{\odot}$.  Unfortunately, we can not use the P--L
relation, since IRAS had insufficient temporal coverage to reliably
constrain periods. Rather, we approximate the source density by an
axisymmetric distribution at $R_0 = 8$ kpc and choose the average
luminosity which maximizes the likelihood function.  The results for
different number of radial terms are shown in Figure
\ref{fig:luminosity}.
For ten terms, the maximum likelihood of this axisymmetric density is
achieved when $L \approx 3000 L_{\odot}$.  We adopt $\overline L =
3500 L_{\odot}$ which is the low end of published results and
interpret our statistical analysis as a consistency check.

\section{Computational Notes} 
 
Likelihood maximization is the rate limiting step in inferring the
surface density from a source catalog. The cost of computing the
likelihood is proportional to the sample size so analyses of very
large data sets will be technically challenging. Our ``workhorse''
algorithm for locating the maximum of the likelihood function is the
conjugate gradient method which is thoroughly discussed in the
literature (e.g. Press et al. 1988).  We have adopted an
implementation by Shanno \& Phua (1976, CONMIN). The algorithm has
good convergence properties, but requires a good initial
approximation. Near the expected quadratic maximum the convergence
should be extremely rapid. 

However, the likelihood function may have a large number of extrema,
limiting the use of the standard downhill technique. In such cases,
the Simulated Annealing (SA) algorithm (Metropolis et al. 1953; Otten
\& van Ginneken 1989) has the advantage. It places no restrictions on
continuity and easily incorporates arbitrary boundary conditions and
constraints. Adaptive Simulated Annealing (ASA, Ingber 1989)---a
faster version of the SA algorithm---proved to be effective in
narrowing the domain of the search to the comparatively small region
in parameter space. However, in the vicinity of the extremum it
converges slowly.  

The complementary features of the two techniques, suggest the
following two-step hybrid scheme:
\begin{enumerate}
\item Use a directed search algorithm (ASA) to isolate the global maximum.
Although SA class of algorithms converge slowly, there is a probabilistic
guarantee of convergence: the probability of finding the maximum is
inversely proportional to the total number of iterations to some power
(e.g. Shu \& Hartley 1987; Ingber 1993).
\item After either a limiting number of steps or a significant drop off in
convergence, use the current ASA solution as input to conjugate gradient
scheme. This is motivated by our expectation that the true maximum of the
likelihood function will be a quadratic form in the unknown variables.
\end{enumerate}
This sequence can be repeated again, in case if Step 2 fails to find a
well-defined maximum. The scheme is difficult to analyze but appears
to work well in practice and is potentially useful for large parameter
space and complex geometry (boundary conditions, irregular likelihood
function) cases.

The entire computation time scales as the number of coefficients M
(total number of $A_{ij}$ and $B_{ij}$ in the sum in eq. \ref{eq:d15})
and the sample size N: $N (2 M + 1)$. Computation of the Hessian
matrix requires CPU time proportional to $M^3 N$. For a large $M$,
this is the bottleneck.  However, the algorithm is straightforwardly
parallelized by partitioning the data.

\newpage

\figcaption{\label{fig:sp_distr.1} The sample of 5,500 IRAS PSC variables
(dots). The Sun is located at $X=-8$, $Y=0$.
The data from the shaded sectors are eliminated from the analysis.
The circle shows the distance in the plane where an AGB star
($L = 3500 L_{\odot}$) can be detected.}

\figcaption{\label{fig:sp_distr.2} The same sample projected on the X-Z plane.
All the data are inside the region bounded by two solid lines which are
solutions of the equation (\protect{\ref{eq:b1}}).}

\figcaption{\label{fig:test} The amplitude of harmonic coefficients as
functions of the position angle of the bar.  Open triangles: $i=1,
j=0$; open squares: $i=1, j=2$; filled triangles: $i=2, j=0$; filled
squares: $i=2, j=2$. The symbols are slightly offset along the
$x$-axis for clarity.}

\figcaption{\label{fig:dens_all} The reconstructed density profiles. Ten
equally spaced contours between $10$\% and $100$\% of peak value are shown
in each panel.}

\figcaption{\label{fig:dens_even} The reconstructed density profile obtained
with assumption of bisymmetric source density. There are $10$ contours between
$10$\% and $100$\% of peak value in each panel.}

\figcaption{\label{fig:levels} Isophotal fits to the reconstructed source
density: surface density $C$ normalized to its central value (left scale,
solid line) and axis ratio $a:b$ (right scale, dashed line)
versus semimajor axis.}

\figcaption{\label{fig:angles} The position angle $\phi$ in degrees (left scale,
solid line) and eccentricity of ellipses (right scale, dashed line) versus
semimajor axis.}

\figcaption{\label{fig:dwek} The distribution of the scale lengths $r_0$
in $5000$ realizations of the source density (histogram).  The best fit normal
distribution is shown (solid curve) with mean and rms value as
labeled.}

\figcaption{\label{fig:baadetau} The ratio of optical depths toward Baade's window
obtained with non-axisymmetric (bar) and axisymmetric disk models
as the function of the position angle of the bar, $\phi_0$.
Solid line -- ``maximal disk'', dashed line -- ``canonical disk'' (see text).
}

\figcaption{\label{fig:reldiff} Asymmetry in the microlensing optical depth. The
disk is ``maximal''.
Solid line, dashed line and dotted line represent cuts with
$b = 0^{\circ}, 2^{\circ}$ and $4^{\circ}$, correspondingly.  }

\figcaption{\label{fig:meantau} Average optical depth as the function of the
galactic longitude. The lines represent the same latitudes as in
Fig. \protect{\ref{fig:reldiff}}}

\figcaption{\label{fig:luminosity} The values of the likelihood function with
varying luminosity of the sources. The center of the Galaxy is fixed at $8$
kpc. Solid line --- $i_{max}=2$, dotted line --- $i_{max}=6$,
dashed line --- $i_{max}=10$.}

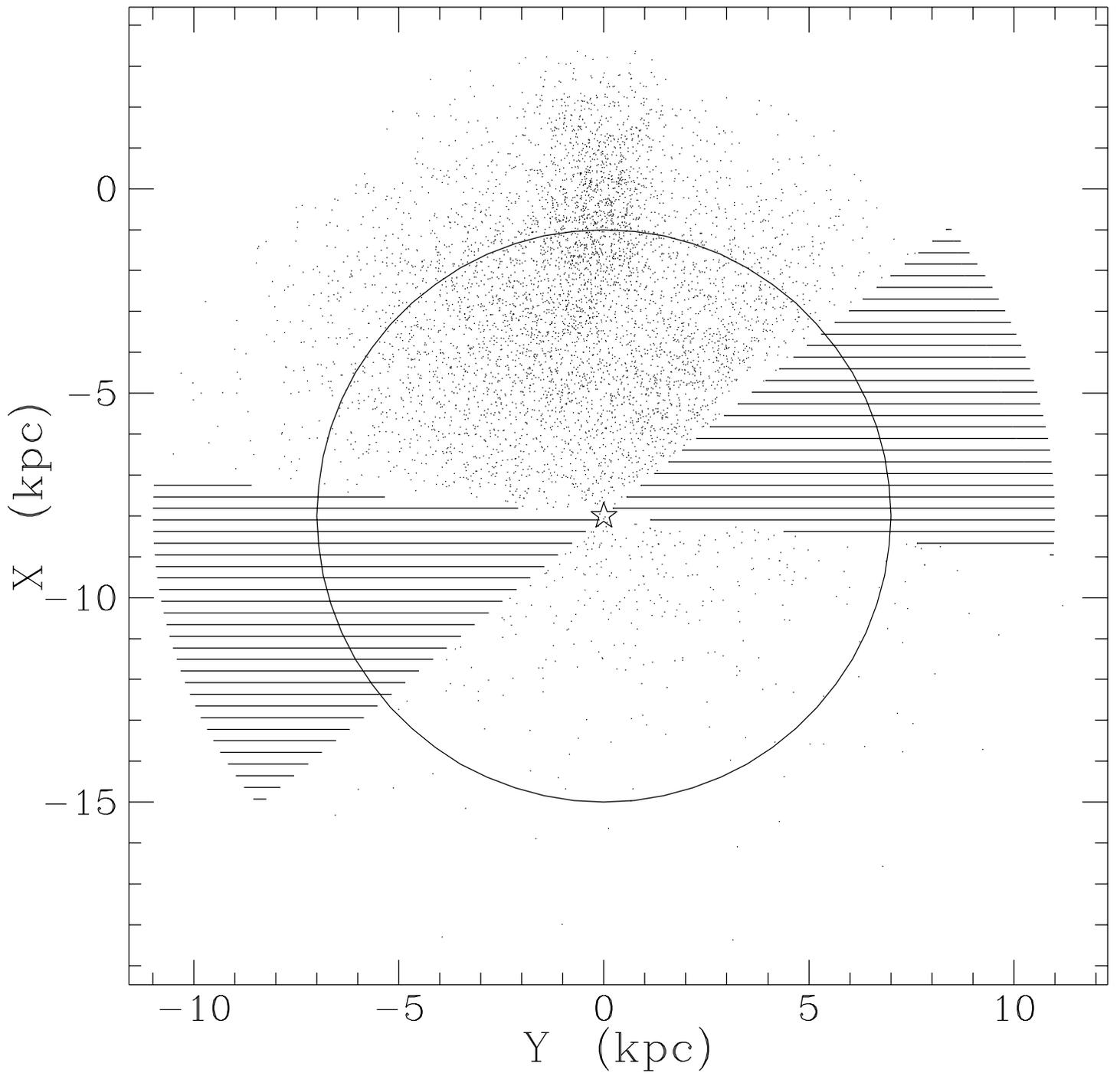
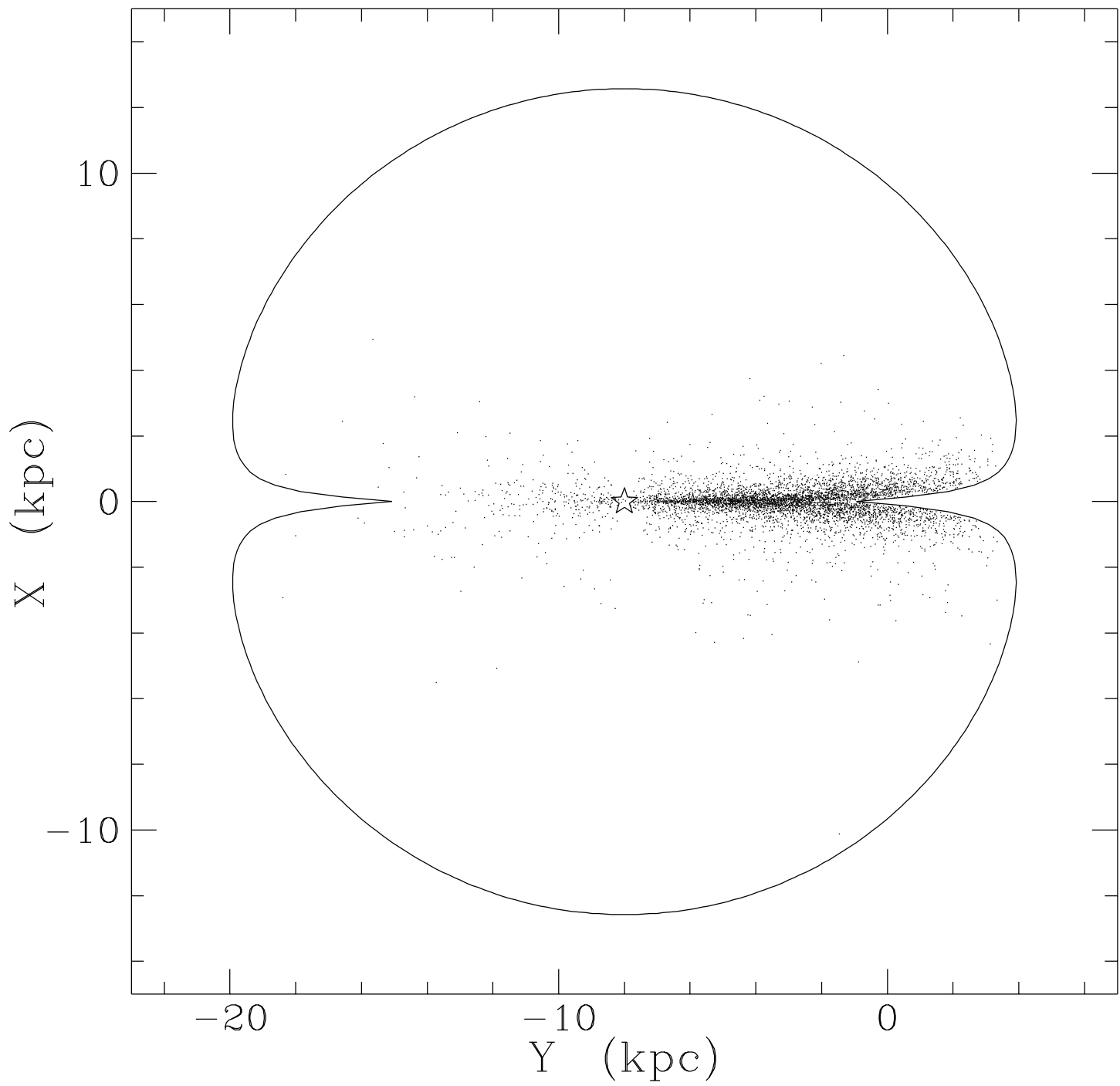
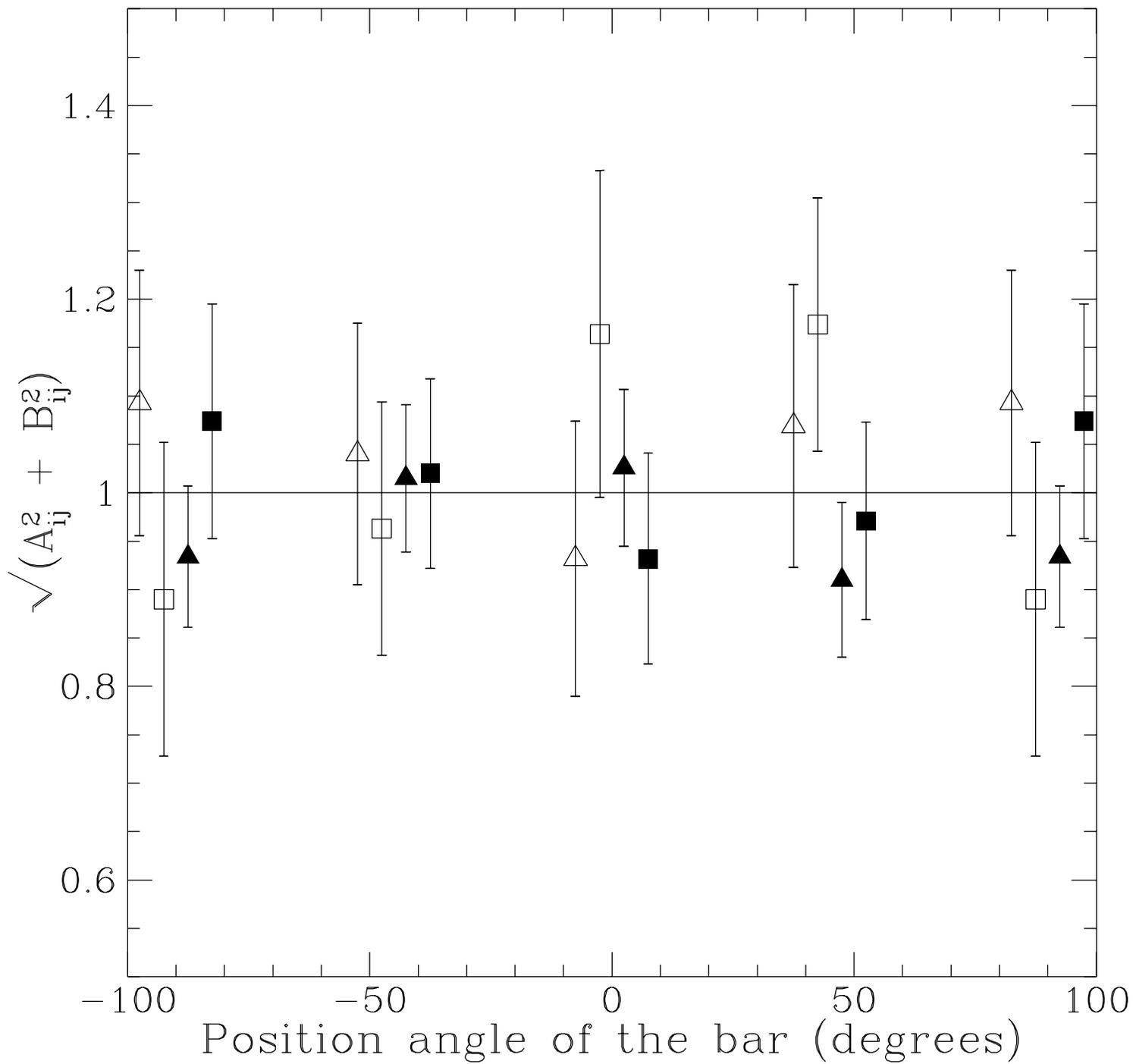
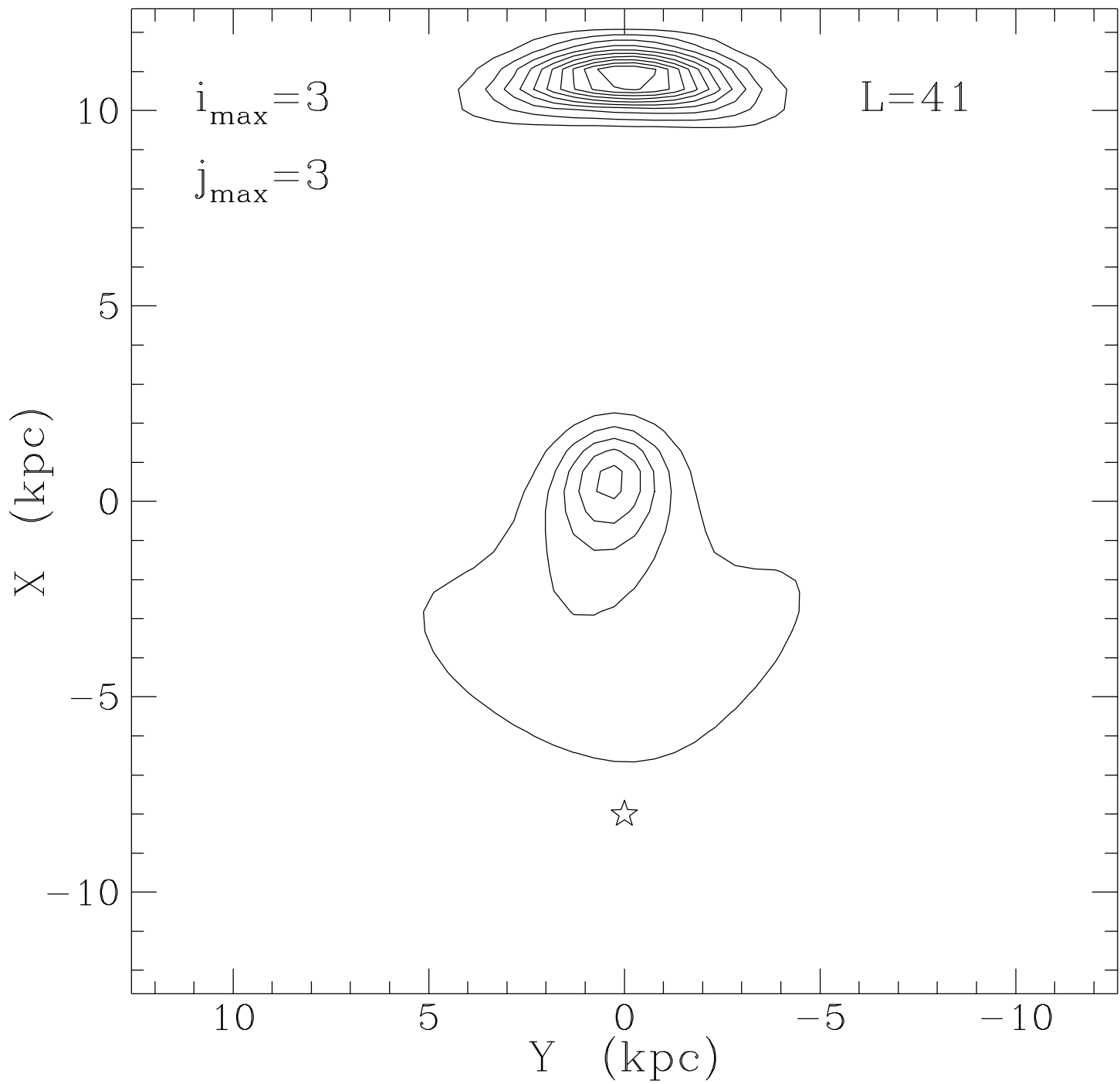
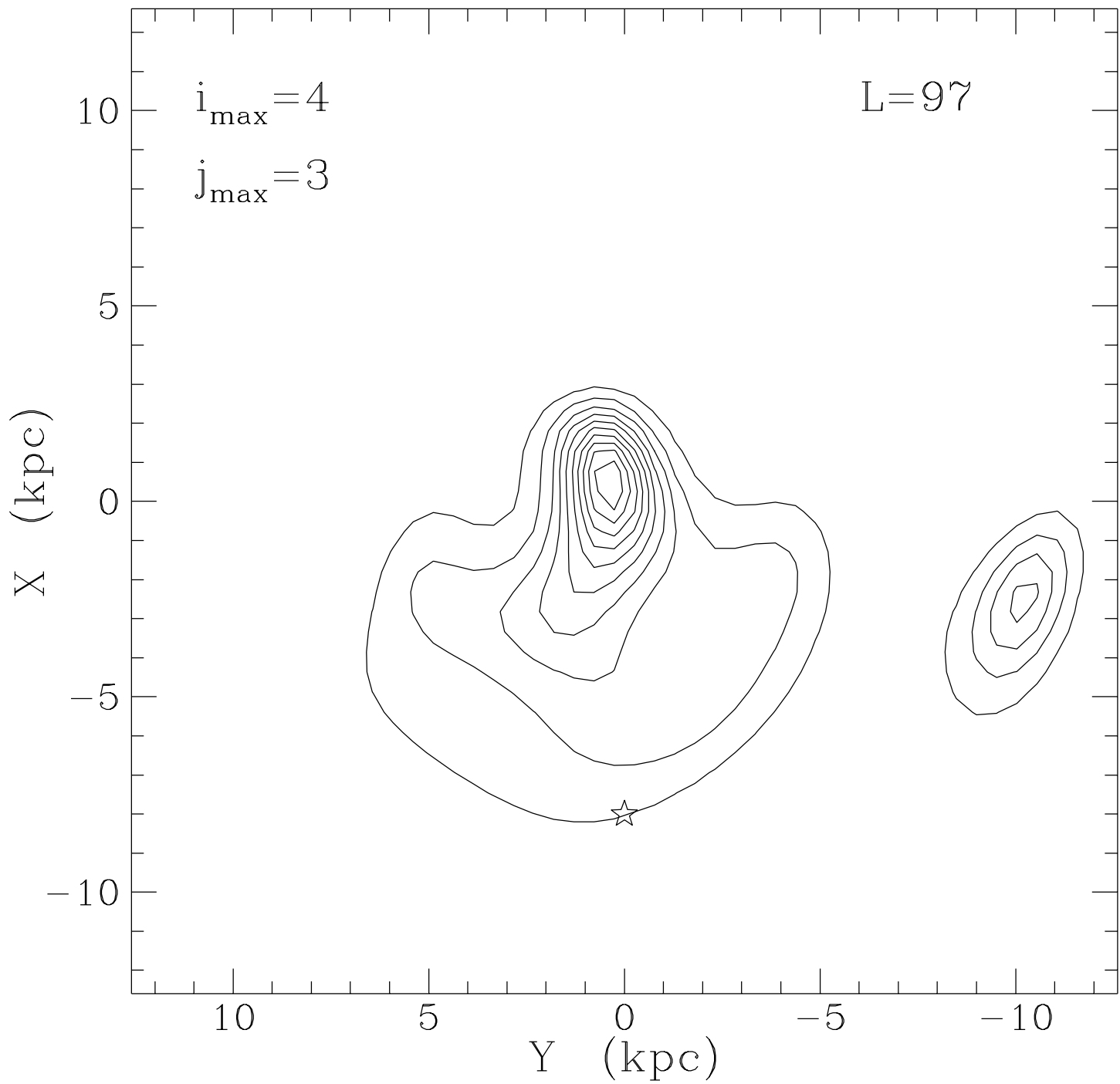
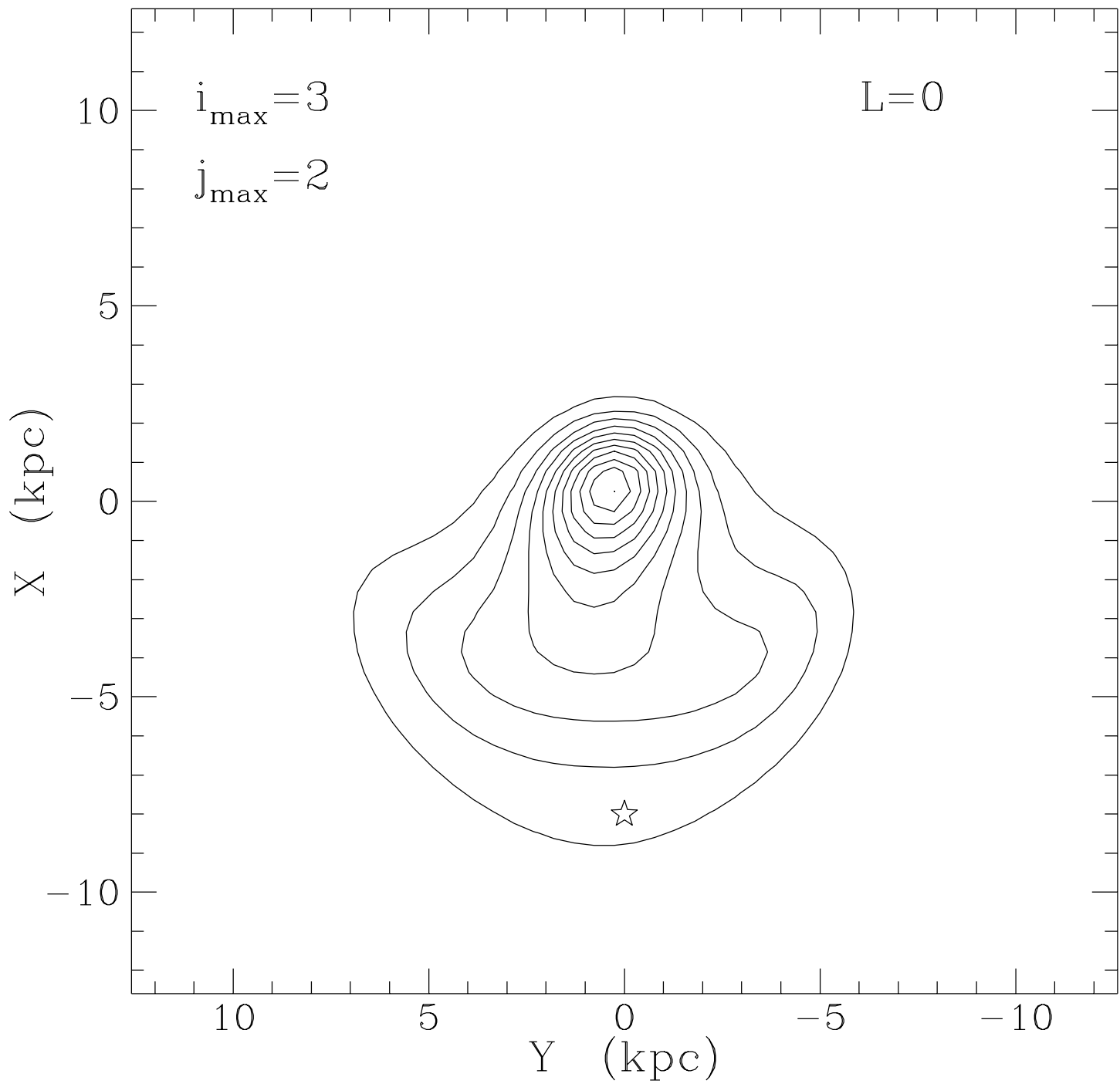
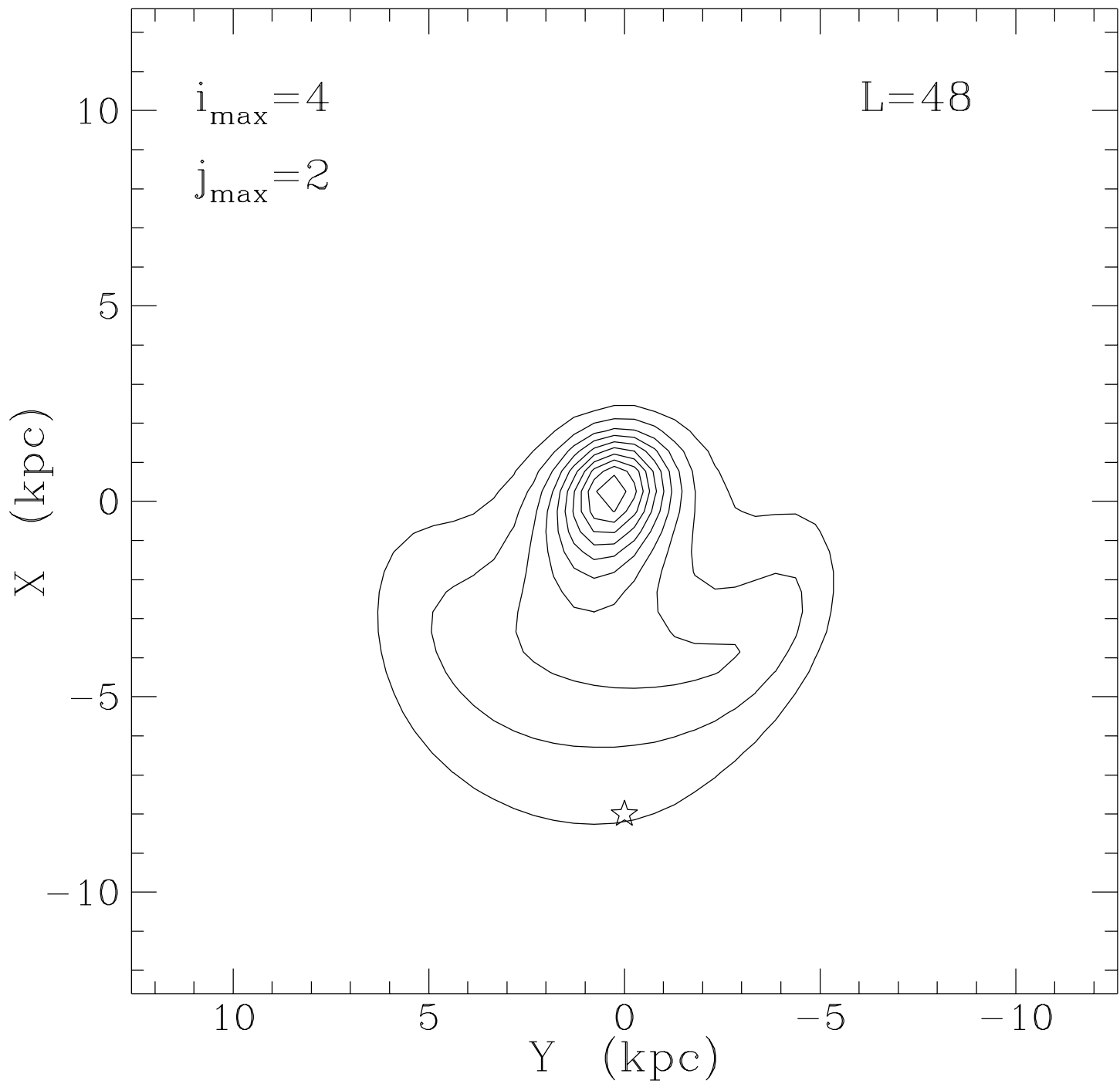
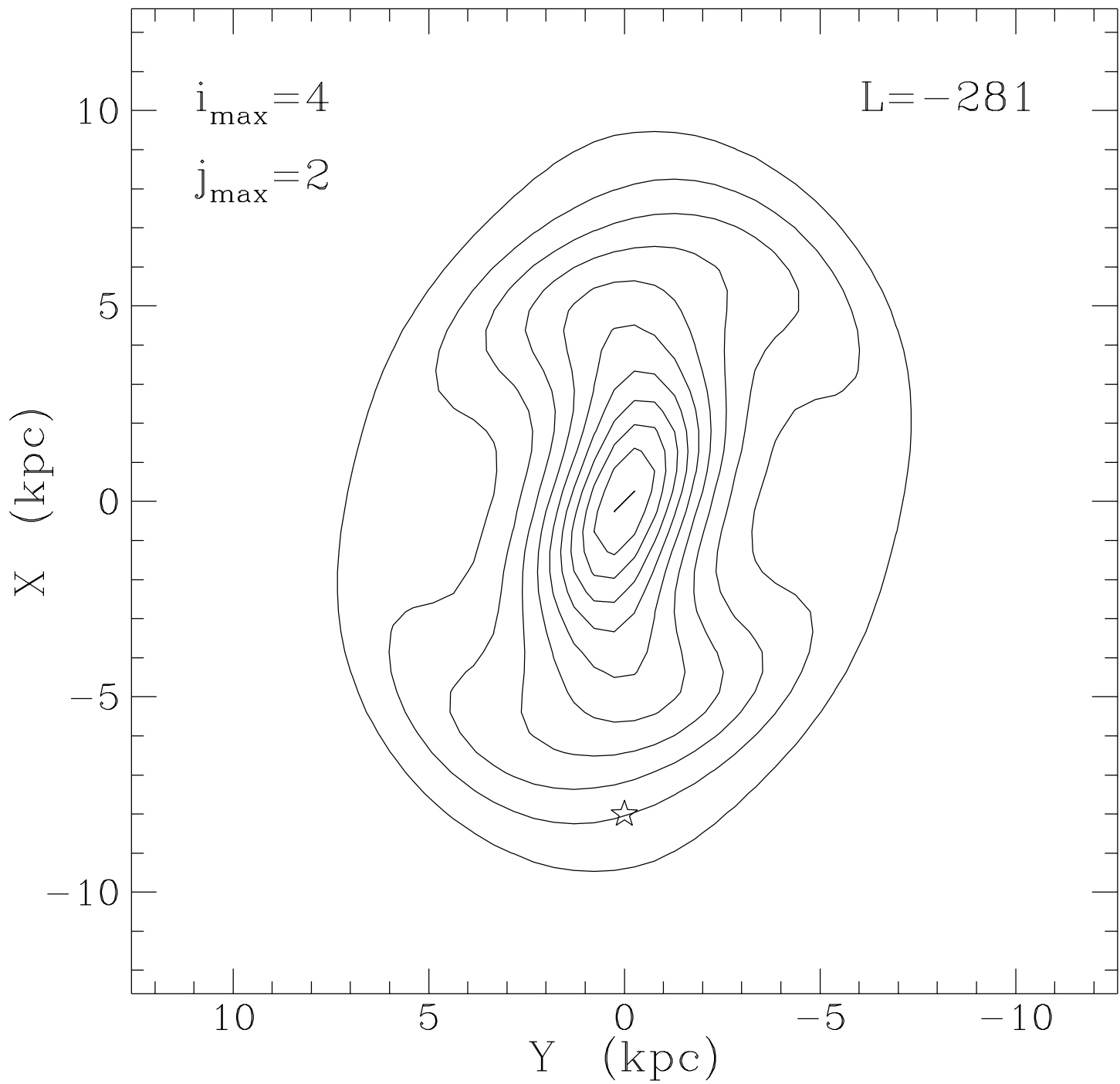
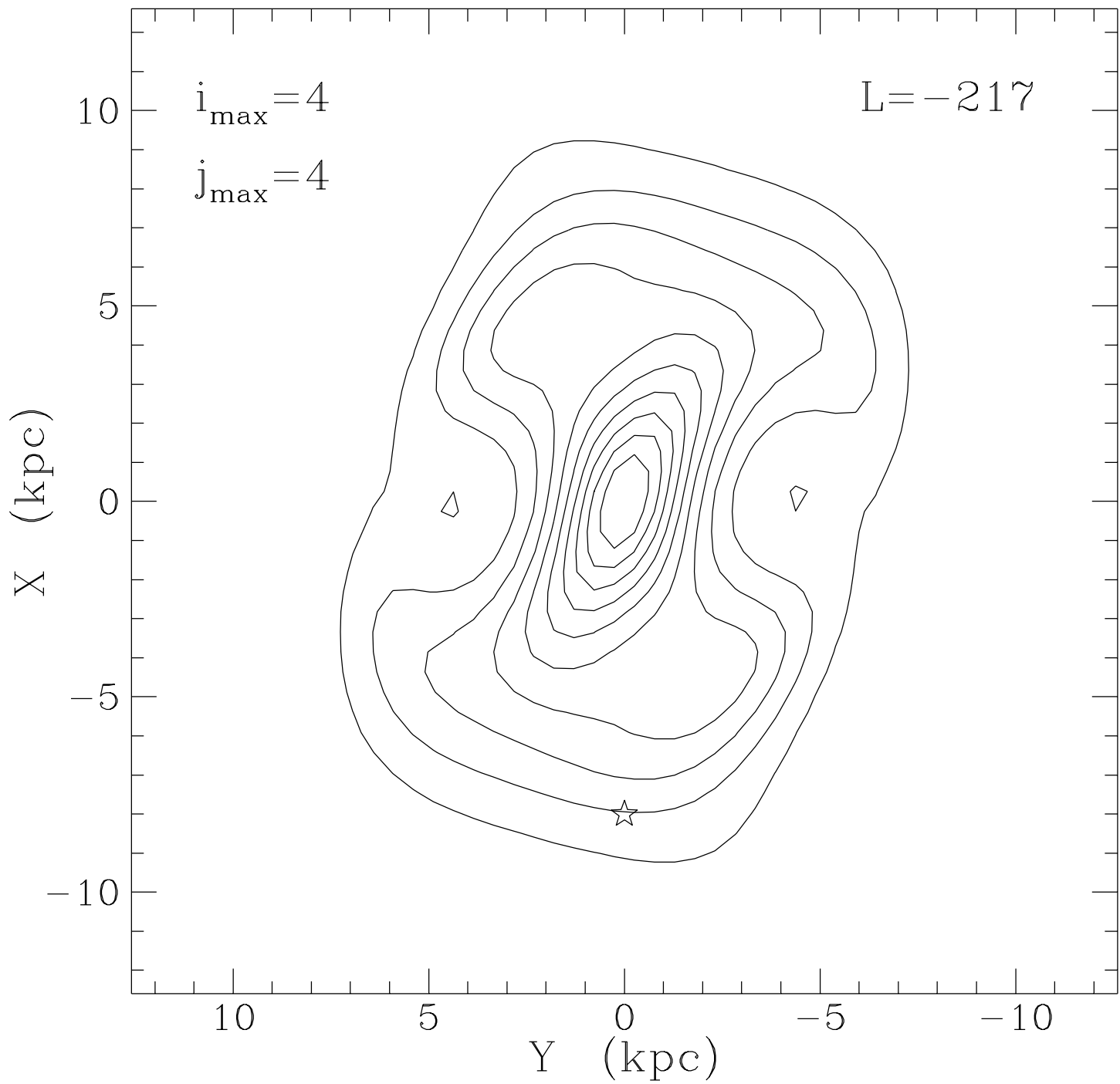
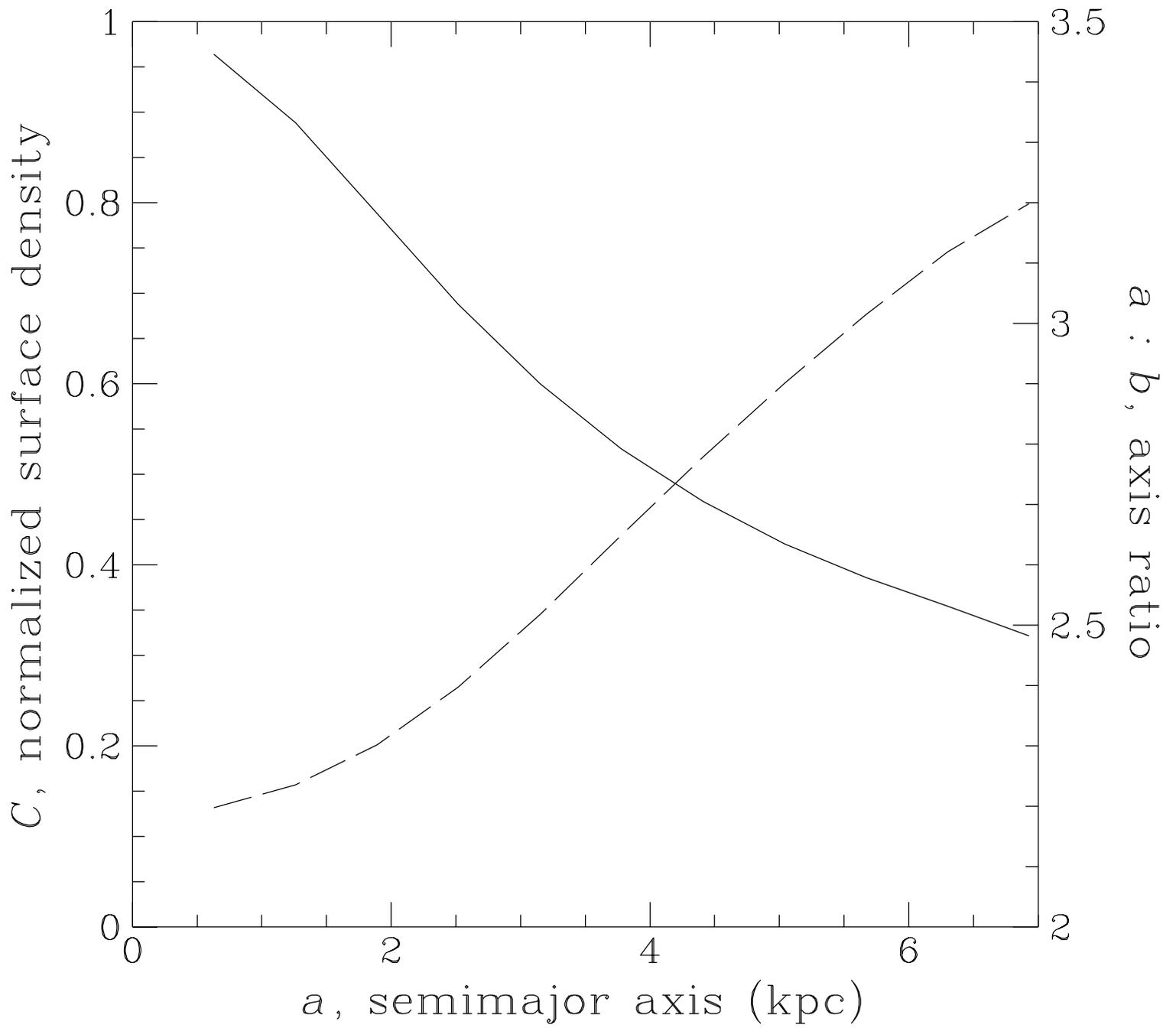
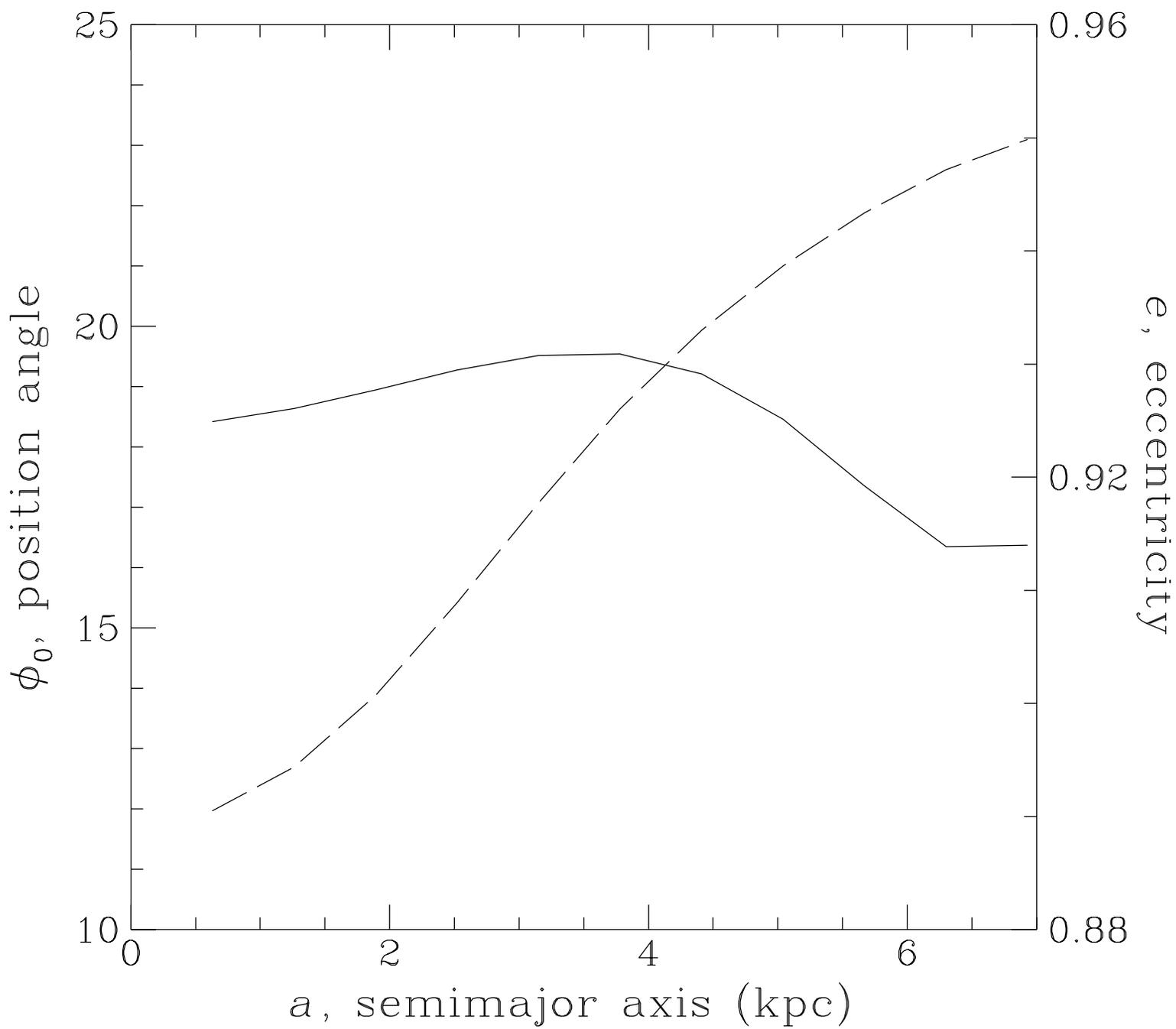
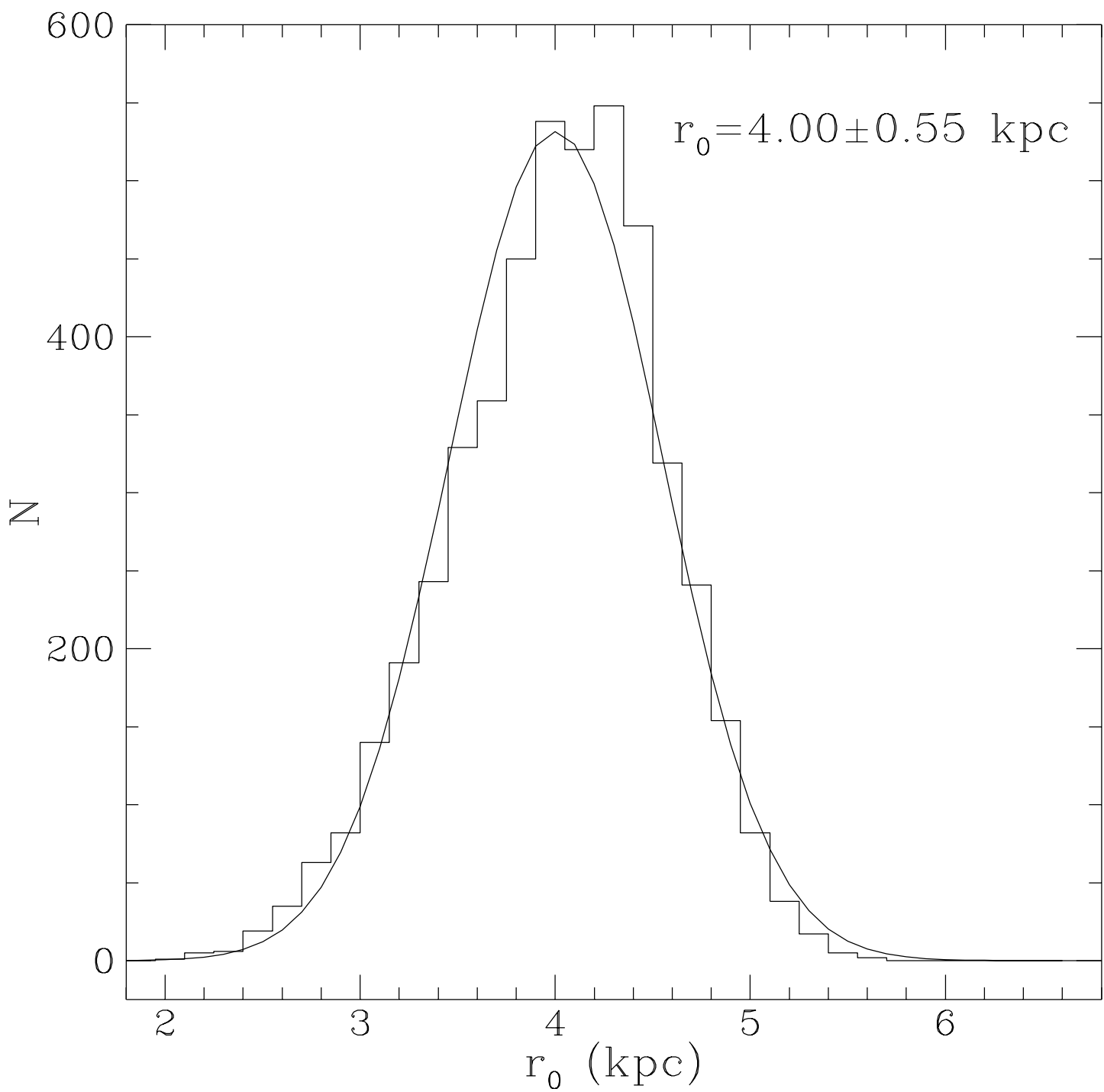
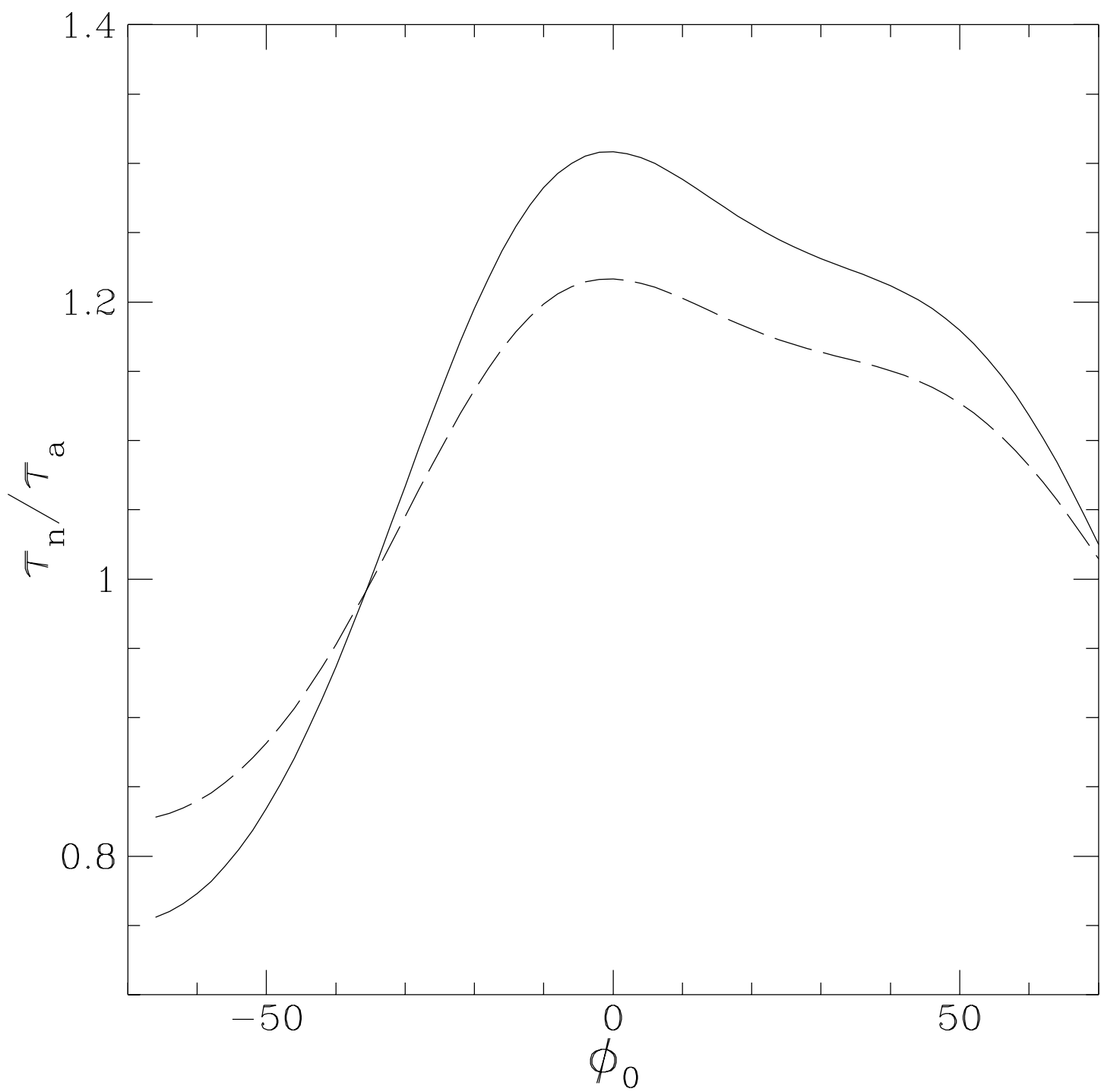
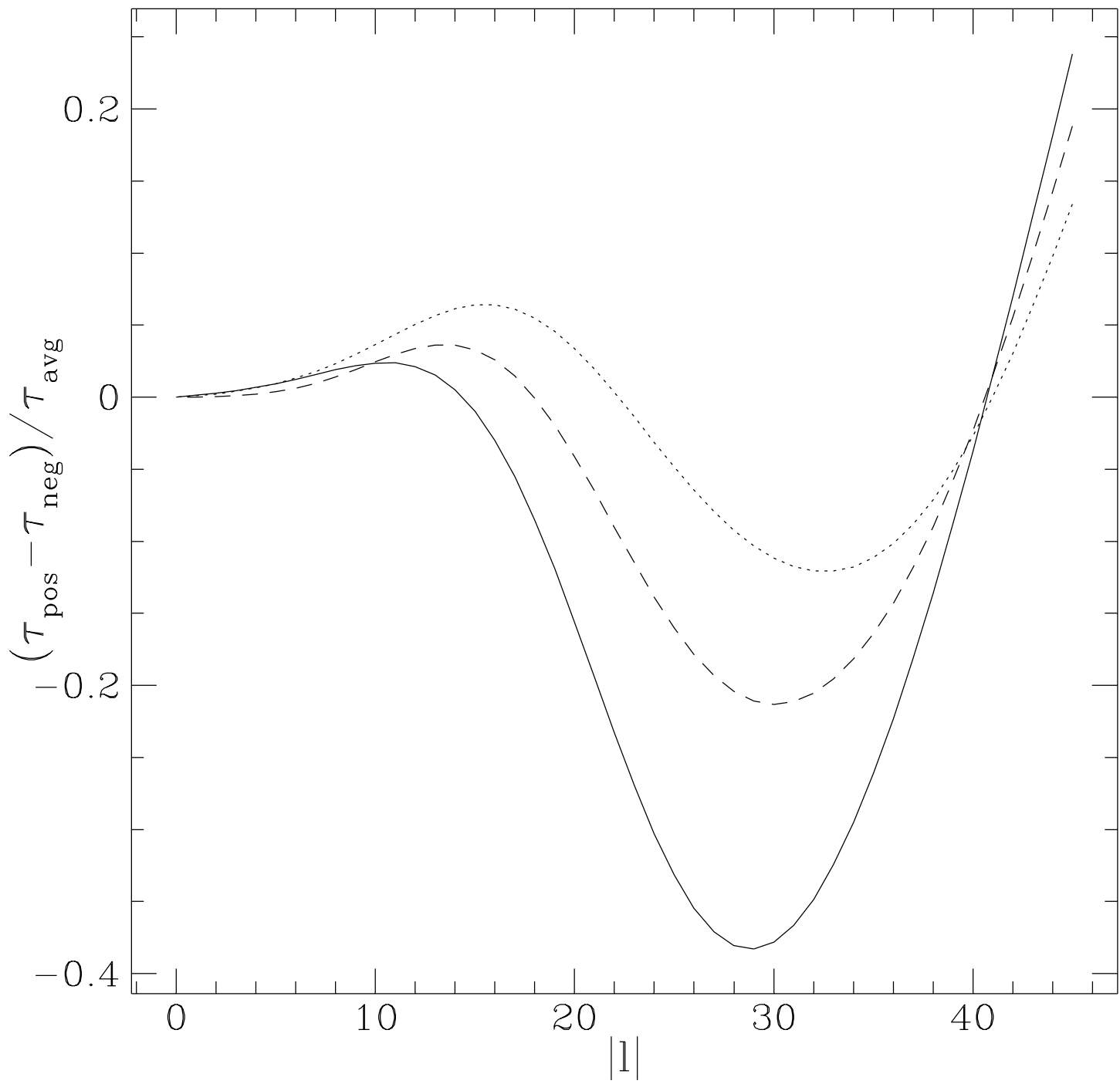
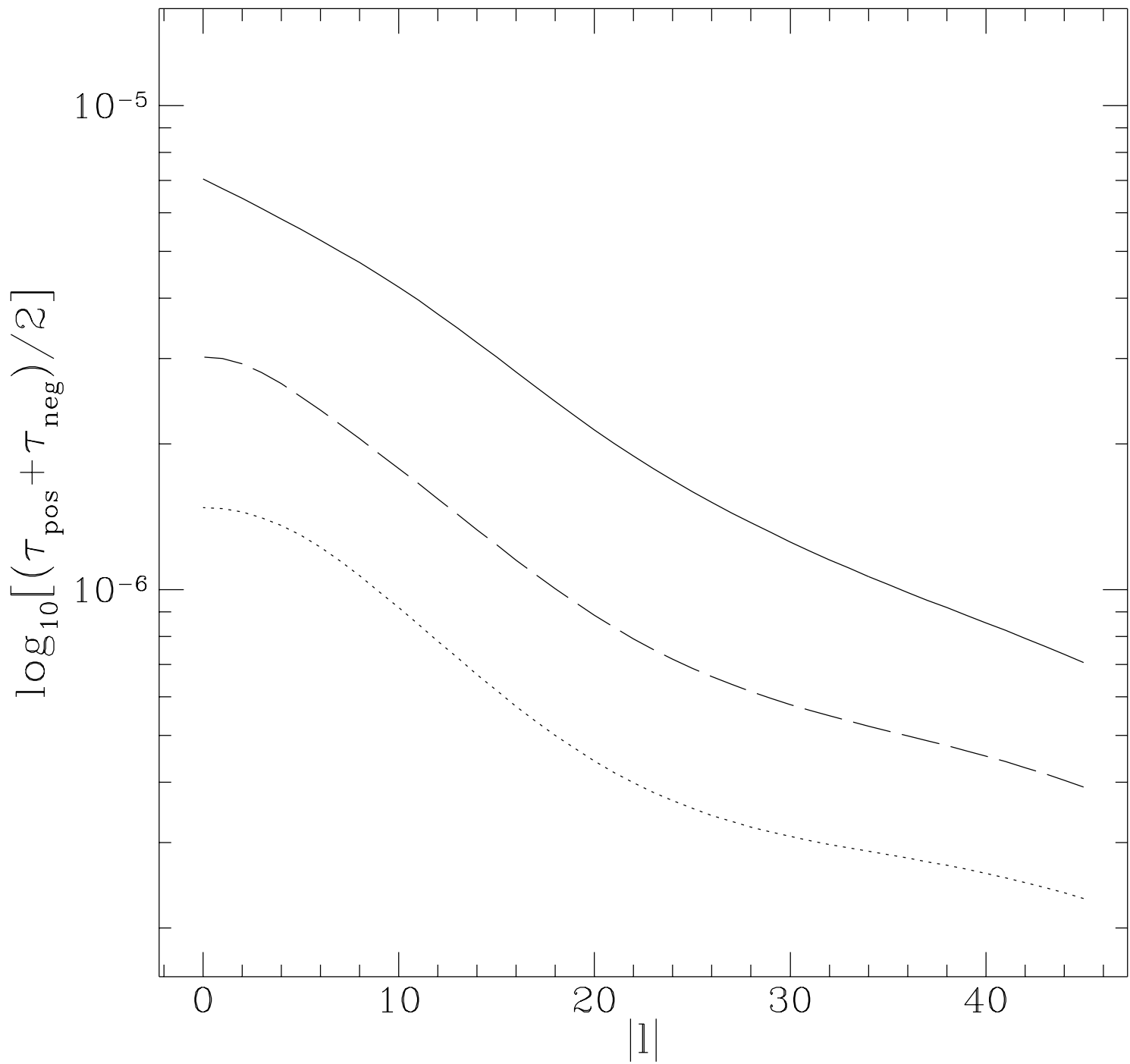
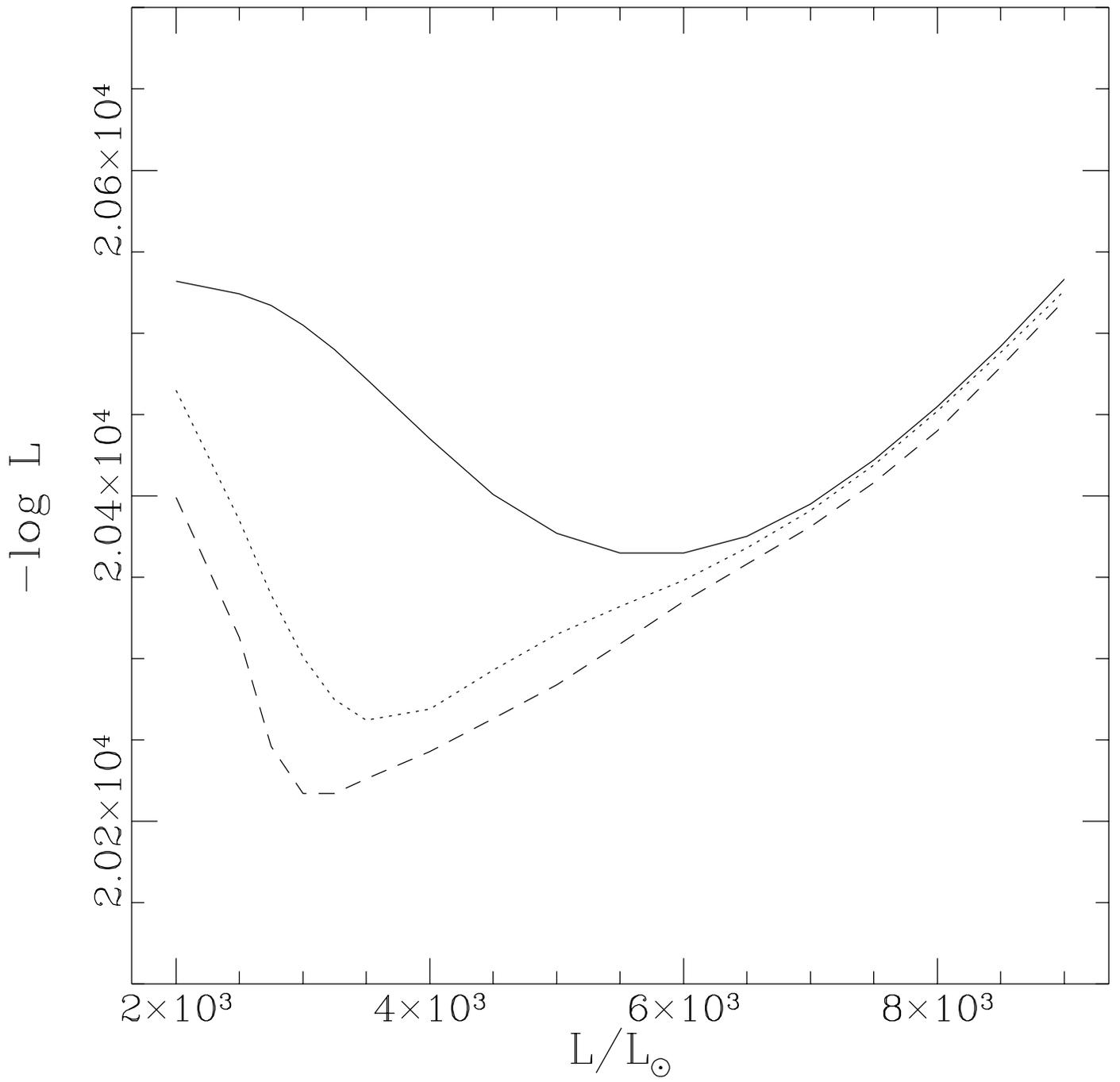
\end{document}